\documentclass[figures]{epl}
\usepackage{amsmath,amssymb,graphicx}

\title{Dynamics and steady state of a vibrated granular binary mixture model}
\shorttitle{Model for a vibrated granular binary mixture}
\author{A. Prados \and J. J. Brey}
\institute{F\'{\i}sice Te\'orica, Universidad de Sevilla, Apartado de Correos
1065, E-41080 Sevilla, Spain }

\begin{document}

\maketitle

\begin{abstract}
A model for the dynamical evolution of a granular binary mixture is analyzed.
This system is submitted to a tapping procedure, similarly to what is done in
real experiments. In the weak vibration limit, an effective dynamics for the
tapping process is derived, and the steady state probability distribution is
analytically found. The steady probability does not depend on the details of
the configuration, but only on the number of particles of each of the two
species. Depending on the values of their fugacities, the system can be either
almost full of small or big particles, i.e., segregation effects are present.
\end{abstract}

Granular systems have attracted the attention of physicists in recent years. A
review of some of the basic phenomenological features of granular matter can be
found in Refs.~\cite{JNyB96,Ka98}. Many of these behaviours, like compaction
\cite{KFLJyN95} or segregation \cite{Wi76,RSPyS87} are far from being well
understood. In statistical physics, simple models are often used as a first
approximation to many different, complex,  problems. Thus, recently, several
Ising or particle-hole models \cite{BPyS99,BKLyS00,LyD01,ByM01,PyB02,BFyS02}
have been used to analyze compaction processes in dense granular media. Also,
the parking lot model has been employed in the same context\cite{BKyN96}.

One of the most interesting physical questions in the physics of dense granular
media is the description of the steady state eventually reached by a system,
when it is externally perturbed. Thermodynamics is not directly applicable to
powders. Nevertheless, some years ago, Edwards and coworkers \cite{EyO89} made
the hypothesis that the steady state of an externally perturbed granular system
can be described by an extension of the usual statistical mechanics concepts.
The central point is the ergodic hypothesis for externally perturbed powders:
in the steady state, all the mechanically stable (metastable) configurations of
a granular assembly occupying the same volume have the same probability.

When a granular mixture is shaken, their components often tend to separate,
leading to the well-known phenomenon of segregation \cite{JNyB96,Ka98}.
Usually, when a bidisperse granular system is vibrated, larger particles go to
the top, the so-called Brazil Nut effect (BNE) \cite{Wi76,RSPyS87,MLNyJ01}.
Very recently, it has been found that, as the mass density of the big grains is
increased as compared with that of the small grains, this phenomenon can be
inverted: small grains go to the top while large grains move to the bottom of
the container \cite{HQyL01,BEKyR03}. This is known as the reverse Brazil Nut
effect (RBNE).

One of the main purposes of this paper is trying to understand some aspects of
the dynamical behaviour of binary granular mixtures by means of the analysis of
a simple, analytically tractable, model. In particular, we will be interested
in the study of the steady state reached by the system in the long time limit.
Its relationship with Edwards' theory will be discussed, and also the
appearance of both the BNE and RBNE behaviours.

Let us consider an horizontal section of a real granular binary mixture, near
the bottom of its container. During the free evolution of the system, i.e.,
only under the action of gravity, particles can only go down, as long as there
is enough empty space in their surroundings.  The total density of particles in
the layer grows until the hard-core interaction prevents more movements of
particles, and a \textit{metastable} (mechanically stable) configuration is
reached. On the other hand, when the system is submitted to vertical vibration,
particles can go up, decreasing the density in the layer. In both processes,
big particles will need more free space in their surroundings than small
particles to be adsorbed on or desorbed from the layer.

We introduce a one-dimensional lattice of $N$ sites, which can be either empty
(occupied by a hole), occupied by a small particle A, or occupied by a big
particle B. Variables $n_i$ and $m_i$ are defined as follows. If site $i$ is
occupied by a particle A, $n_i=0$, otherwise $n_i=1$. If site $i$ is occupied
by a particle B, then $m_i=0$, if not, $m_i=1$. Then, an empty site is given by
$n_i=m_i=1$. The numbers $1-n_i$ and $1-m_i$ are the occupation numbers of
particles A and B, respectively. The specification of $\bm{n}=\left\{ n_i
\right\}$ and $\bm{m}=\left\{ m_i \right\}$ characterizes a configuration of
the lattice. The dynamics of the system is assumed to be a Markov process
defined by the following master equation for the probability
$p(\bm{n},\bm{m},t)$ for finding the system in the configuration $\left\{
\bm{n},\bm{m} \right\}$ at time $t$,
\begin{eqnarray}
\label{1}
 \partial_t p(\bm{n},\bm{m},t) & = &
 \sum_i \left[ W_i(\bm{n},\bm{m}|R_i\bm{n},\bm{m}) p(R_i\bm{n},\bm{m},t)-
 W_i(R_i\bm{n},\bm{m}|\bm{n},\bm{m}) p(\bm{n},\bm{m},t) \right] \nonumber \\
 & & +
 \sum_i \left[ W_i(\bm{n},\bm{m}|\bm{n},R_i\bm{m}) p(\bm{n},R_i\bm{m},t)-
 W_i(\bm{n},R_i\bm{m}|\bm{n},\bm{m}) p(\bm{n},\bm{m},t) \right] \, .
\end{eqnarray}
Here, $R_i\bm{m}=\{ \ldots, m_{i-1},1-m_i,m_{i+1},\ldots \}$ and $R_i\bm{n}=\{
\ldots,n_{i-1},1-n_i,n_{i+1},\ldots \}$. Thus, the possible events are the
adsorption of an A or B particle on an empty site, and their desorption from an
occupied site. In order to model the hard-core interactions, a facilitated
dynamics \cite{FyA84} is considered. Small particles A need, to be adsorbed on
or desorbed from a site of the lattice, at least one of its two nearest
neighbour sites being empty. Namely,
\begin{equation}
\label{2}
 W_i(R_i\bm{n},\bm{m}|\bm{n},\bm{m})=\left[ \frac{\alpha_a}{2} n_i m_i+
 \frac{\alpha_d}{2} (1-n_i) m_i \right] \left( n_{i-1}m_{i-1}+n_{i+1}m_{i+1}
 \right) \, ,
\end{equation}
where $\alpha_a$ and $\alpha_d$ are the characteristic rates for the attempts
of adsorption  and desorption of particles A, respectively. On the other hand,
big particles B need both of its nearest neighbours empty when adsorbing on a
site or desorbing from it,
\begin{equation}
\label{3}
 W_i(\bm{n},R_i\bm{m}|\bm{n},\bm{m})=\left[ \beta_a n_i m_i + \beta_d n_i
 (1-m_i) \right] n_{i-1} m_{i-1} n_{i+1} m_{i+1} \, ,
\end{equation}
$\beta_a$ and $\beta_d$ being the rates for the attempts of particles B to
adsorb on and desorb from the lattice, respectively. If particles $B$ are
eliminated, i.e., $m_i=1$ for all $i$ and $\beta_a=0$, our system reduces to a
model for a single type of particles \cite{BPyS99,PByS00}.

We are interested in the particularization of the above general dynamics for
the tapping process used in the laboratory to vibrate real granular systems
\cite{KFLJyN95}. It will be modelled as follows, analogously to the procedure
introduced for a monodisperse system \cite{BPyS99,LyD01}. Starting from an
empty lattice, the system is allowed to relax with $\alpha_d=\beta_d=0$, i.e.,
only adsorption events are allowed. The system evolves until it reaches a
metastable configuration, in which no more adsorptions are possible. The
metastable configuration so obtained will be the initial state for the tapping
process and will be referred to as state $r=0$. Next, we describe the vibration
process by making $\alpha_a=\beta_a=0$, particles can only desorb from the
lattice, for a given time interval $t_0$. This pulse takes, in general, the
system out from the metastable configuration. Afterwards, the system relaxes
again with $\alpha_d=\beta_d=0$ until it gets stuck in a new metastable
configuration $r=1$. By repeating this sequence, a series of metastable states
$r=2,3,4,\ldots$ is generated. The index $r$ indicates the number of taps made
before that metastable configuration has been reached. Notice that the
non-conservation of the number of particles in the lattice  tries to mimic what
actually occurs in a deep layer of a vibrated granular system.

The metastable configurations are characterized as follows: (i) all the holes
are isolated, (ii) two big particles are separated by a domain of sites of
length $l\geq 1$ and, moreover, that domain cannot be completely full of
particles A. Thus, the number of holes verifies $N_H=N-N_A-N_B>N_B$. The number
of metastable configurations $\Omega^{(N)}_{N_A,N_B}$ for a lattice of $N$
sites with $N_A$ particles A and $N_B$ particles B is \cite{PyB03b}
\begin{equation}
\label{4}
 \Omega^{(N)}_{N_A,N_B}=\frac{(N-N_A-N_B-1)!(N_A+2N_B+1)!}
                             {N_B!(N-N_A-2N_B-1)!(2N_A+2N_B-N+1)!(N-N_A)!}
                             \, .
\end{equation}
In the large $N$ limit, there will be well defined densities of particles A and
B, $\rho_A\equiv N_A/N$ and $\rho_B\equiv N_B/N$, and the number of states
$\Omega^{(N)}_{N_A N_B}$ is exponentially large. Thus,  $\ln\Omega^{(N)}_{N_A
N_B}/N$ is only a function of $\rho_A$ and $\rho_B$ and, therefore, independent
of $N$ itself, i. e., the ``microcanonical'' entropy
$S\equiv\ln\Omega^{(N)}_{N_A N_B}$ is an extensive property.

The chain of metastable configurations obtained by the tapping procedure
defines a Markov process: no information is needed from the metastable state
$r-1$ in order to compute the probability of going from configuration $r$ to
$r+1$. This is due to the free relaxation occurring between every two vibration
pulses.  Therefore, it is tempting to try to identify the transition
probabilities $W_{\ab{ef}}(\bm{n}^\prime,\bm{m}^\prime|\bm{n},\bm{m})$ from the
initial metastable configuration $\{ \bm{n},\bm{m}\}$ to the final one $\{
\bm{n}^\prime , \bm{m}^\prime \}$ in a single tap. This can be done if
$\alpha_d t_0\ll 1$, $\beta_d t_0\ll 1$, i.e., the probability for the
desorption events during the vibration is very small. Then, an expansion in
powers of $\alpha_d t_0$ and $\beta_d t_0$ is expected to be useful.

\begin{figure}
\twofigures[scale=0.3]{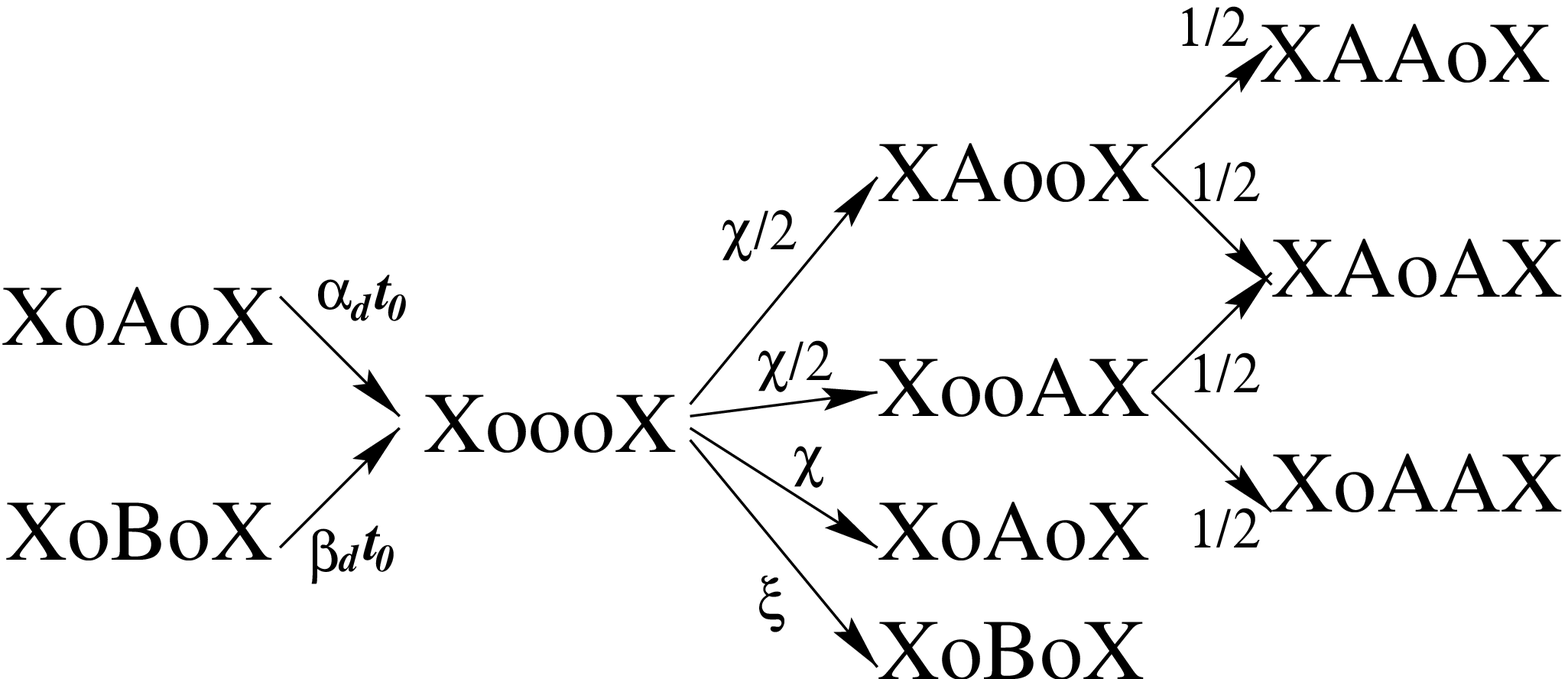}{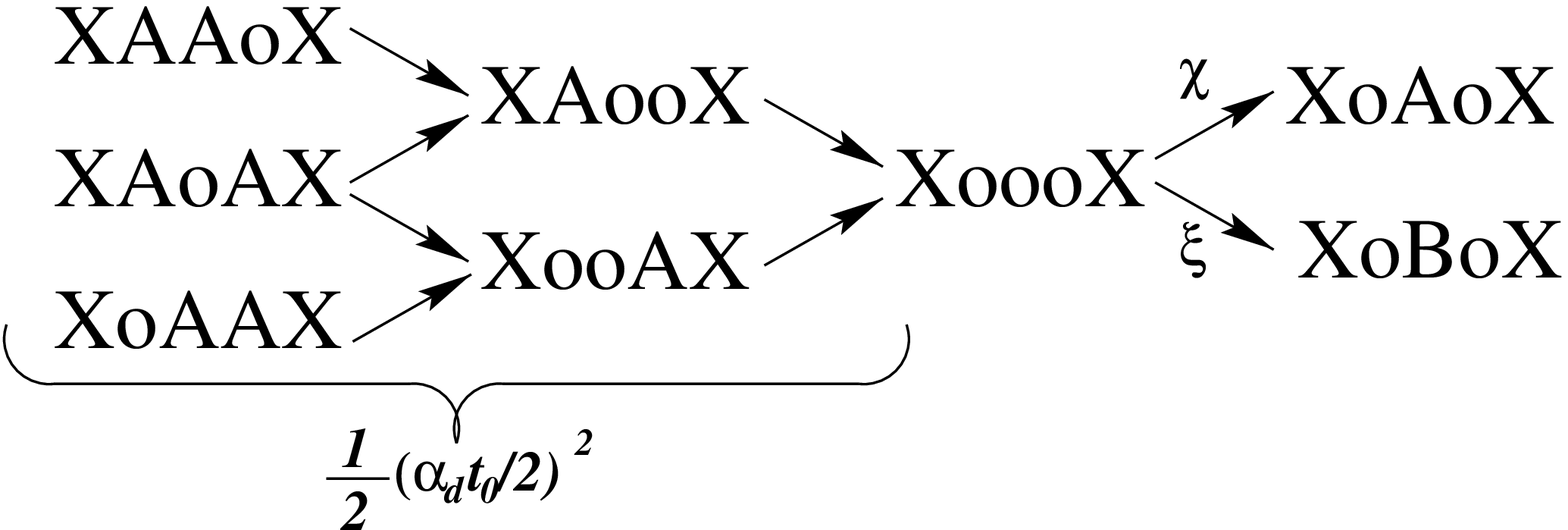}
 \caption{
Desorption of a particle between two empty sites, followed by the events during
the free relaxation necessary to get into a metastable configuration.}
 \label{f1}
 \caption{
Second order processes being the inverse of those first order trajectories
increasing the density of the system. Two desorption events during the pulse of
duration $t_0$ are needed.}
 \label{f2}
\end{figure}

The lowest order is equivalent to consider that each site is affected by only
one rearrangement in a single tap. Due to the facilitated dynamics, only
particles A next to at least one empty site or particles B between two empty
sites can desorb, leading the system to an unstable configuration. Afterwards,
the free relaxation will take the system back to a metastable configuration.
Let us consider a small particle A next to only one hole. It desorbs during the
vibration with probability $\alpha_d t_0/2$. Afterwards, in the free
relaxation, a small particle A is adsorbed in any of the two neighbouring empty
sites with the same probability $1/2$. Therefore, the probability for a
one-site diffusion process of small particles A is $\alpha_d t_0/4$, as shown
in Table \ref{t1}. There, the symbol o represents an empty site, and X stands
for an occupied site, either by a particle A or a particle B. On the other
hand, there are no diffusive processes for particles B up to this order, since
they need both of their nearest neighbours empty in order to desorb. The other
lowest order processes start with the desorption of a particle between two
empty sites. The possible trajectories are shown in Fig.\ \ref{f1}, in which
the probability for each step is given. We have introduced the parameters
$\chi=\alpha_a/(2\alpha_a+\beta_a)$ and $\xi=\beta_a/(2\alpha_a+\beta_a)$. The
transition probabilities for each rearrangement connecting metastable states is
obtained as the product of the probabilities for each of the steps composing
it, and they are summarized in Table \ref{t1}. We do not compute the
probabilities for the transitions leading to a final state identical to the
initial one, since they are not needed when using a master equation description
\cite{vk92}.

To the lowest order, only compaction events take place, since the total number
of particles $N_A+N_B$ always increases. In order to have processes decreasing
the total density, we are led to consider second order terms, similarly to what
happened in the monodisperse case \cite{BPyS99,PyB02}. In particular, we will
analyze whether the inverse processes of those increasing the density to the
lowest order are possible, and which are their transition probabilities. These
are depicted in Fig. \ref{f2}, together with the probability of each elementary
step. The corresponding effective transition probabilities between metastable
configurations are also included in Table \ref{t1}. The Markov process defined
by the effective transition rates of Table \ref{t1} is irreducible, i.e., all
the metastable configurations are connected by a chain of transitions with
nonzero probability. This assures the existence of a unique stationary solution
of the master equation.

\begin{table}
\caption{Transitions between metastable states, in a single tap, with their
corresponding rates.} \label{t1}
\begin{center}
\begin{tabular}{cccc}
 FIRST ORDER &  Initial state  & Final state  &  $W_{\ab{ef}}$  \\
 \hline
 one-site diffusion & XAoX  & XoAX  & $\alpha_d t_0/4$ \\
  & XoAX  & XAoX  & $\alpha_d t_0/4$ \\
  change B into A & XoBoX & XoAoX & $\beta_d t_0 \chi$ \\
  change A into B & XoAoX & XoBoX & $\alpha_d t_0 \xi$ \\
  change B into 2A & XoBoX & XAoAX & $\beta_d t_0 \chi/2$  \\
  & XoBoX & XAAoX or XoAAX & $\beta_d t_0 \chi/4$ \\
  change A into 2A & XoAoX & XAoAX & $\alpha_d t_0 \chi/2$ \\
  & XoAoX & XAAoX or XoAAX & $\alpha_d t_0 \chi/4$ \\
  SECOND ORDER & Initial state  & Final state  &  $W_{\ab{ef}}$ \\
  \hline
  change 2A into B & XAAoX or XoAAX &  XoBoX & $(\alpha_d t_0)^2 \xi/8$  \\
   & XAoAX & XoBoX & $(\alpha_d t_0)^2 \xi/4$ \\
   change 2A into A & XAAoX or XoAAX  &  XoAoX & $(\alpha_d t_0)^2 \chi/8$ \\
   & XAoAX & XoAoX & $(\alpha_d t_0)^2 \chi/4$
\end{tabular}
\end{center}
\end{table}

In order to derive the steady state distribution, we will bet a priori on a
stationary solution $P^{(s)}(\bm{n},\bm{m})$ of the master equation having  the
detailed balance property,
\begin{equation}
\label{9}
 W_{\ab{ef}}(\bm{n},\bm{m}|\bm{n}^\prime,\bm{m}^\prime)
 P^{(s)}(\bm{n}^\prime,\bm{m}^\prime)=
W_{\ab{ef}}(\bm{n}^\prime,\bm{m}^\prime|\bm{n},\bm{m})
 P^{(s)}(\bm{n},\bm{m})   \, .
\end{equation}
Let us consider, first, metastable states without particles B, i.e., $N_B=0$.
All the states with the same number of particles A must have the same
probability in the steady state, since they are connected through (isotropic)
diffusive processes. This means that any metastable state  has a steady
probability which only depends on the values of $N_A$ and $N_B$, because any
metastable configuration can be obtained from a configuration with $N_B=0$
through a chain of diffusive processes and $N_B$ substitutions of particles A
by particles B. Then, application of detailed balance to the processes in Table
\ref{t1} yields, for the steady probability of any metastable state with $N_A$
particles A and $N_B$ particles B,
\begin{equation}
\label{10}
 P^{(s)}_{N_A N_B}=\frac{1}{Z} \gamma_A^{-N_A} \gamma_B^{-N_B} \,
 , \quad
 \gamma_A=\frac{1}{2}\alpha_d t_0 \, , \quad
 \gamma_B=\frac{\alpha_a}{2\beta_a}\beta_d t_0  \, ,
\end{equation}
and $Z$ is a generalized ``partition function'',
\begin{equation}
\label{12}
 Z=\sum_{N_A} \sum_{N_B} \Omega^{(N)}_{N_A N_B} \gamma_A^{-N_A}
 \gamma_B^{-N_B} \, .
\end{equation}
All the states with the same numbers $N_A$ and $N_B$ have the same probability;
this is consistent with Edwards' thermodynamic description of the steady state
of externally perturbed powders \cite{EyO89}. As we are dealing with an open
system having two different kinds of particles, the steady state is described
by two thermodynamic parameters, the ``fugacities'' $\gamma_A^{-1}$ and
$\gamma_B^{-1}$ corresponding to particles A and B, respectively.

In the large $N$ limit, the partition function is readily obtained
by means of the saddle point method, with the result \cite{PyB03b}
\begin{equation}
\label{13}
 \ln \zeta\equiv \frac{1}{N} \ln Z=
 \ln \frac{2 \rho_B+\rho_A}{(2\rho_A+2\rho_B-1) \gamma_A} \, ,
\end{equation}
where $\rho_A$ and $\rho_B$ are the steady densities of particles A and B,
respectively, being $0\leq \rho_A\leq 1$ while $0\leq \rho_B\leq 1/2$, since
there must be at least one hole between every two particles B.  Moreover,
$\rho_H=1-\rho_A-\rho_B\geq \rho_B$. The particle densities are determined by
the saddle point condition
\begin{equation}
    \gamma_A= \frac{(\rho_A+2\rho_B)(1-\rho_A-2\rho_B)(1-\rho_A)}
                        {(1-\rho_A-\rho_B)(2\rho_A+2\rho_B-1)^2} \, ,
                        \quad
    \gamma_B = \frac{(\rho_A+2\rho_B)^2(1-\rho_A-2\rho_B)^2}
                        {\rho_B(1-\rho_A-\rho_B)(2\rho_A+2\rho_B-1)^2}
                        \, ,
    \label{14}
\end{equation}
which give $\rho_A$ and $\rho_B$ in terms of the fugacities. The situation is,
in this sense, analogous to the one found in a recent analysis of a hard-sphere
mixture, where an extension of Edwards' theory with two different values of the
compactivity for each of the species was introduced \cite{NFyC02}. As explained
above, in our model two thermodynamic parameters are present because we are
dealing with an open system. Thus, as in molecular systems, two ``fugacities''
are needed, one for each kind of particles.

\begin{figure}
\twofigures[scale=0.3,angle=-90]{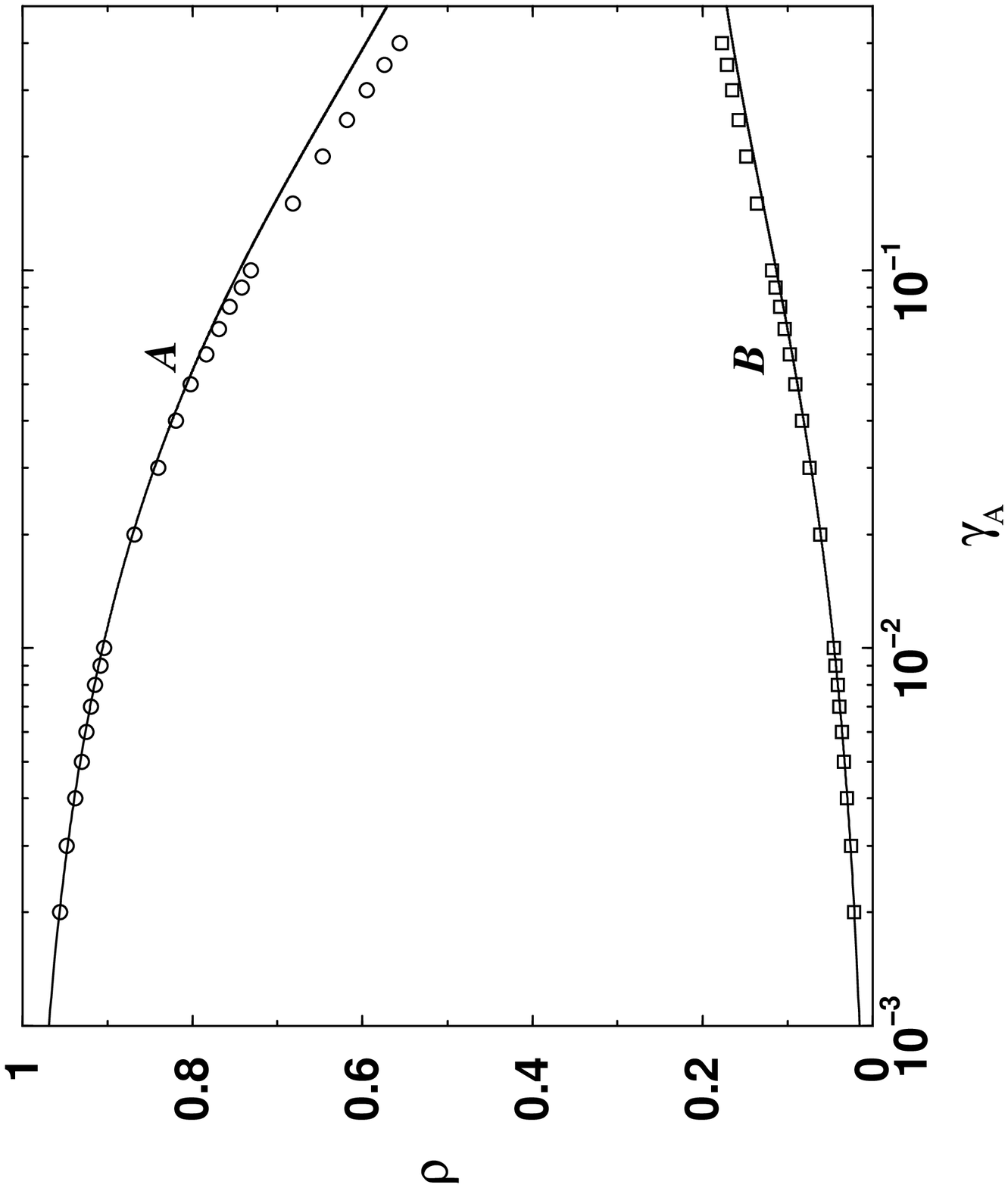}{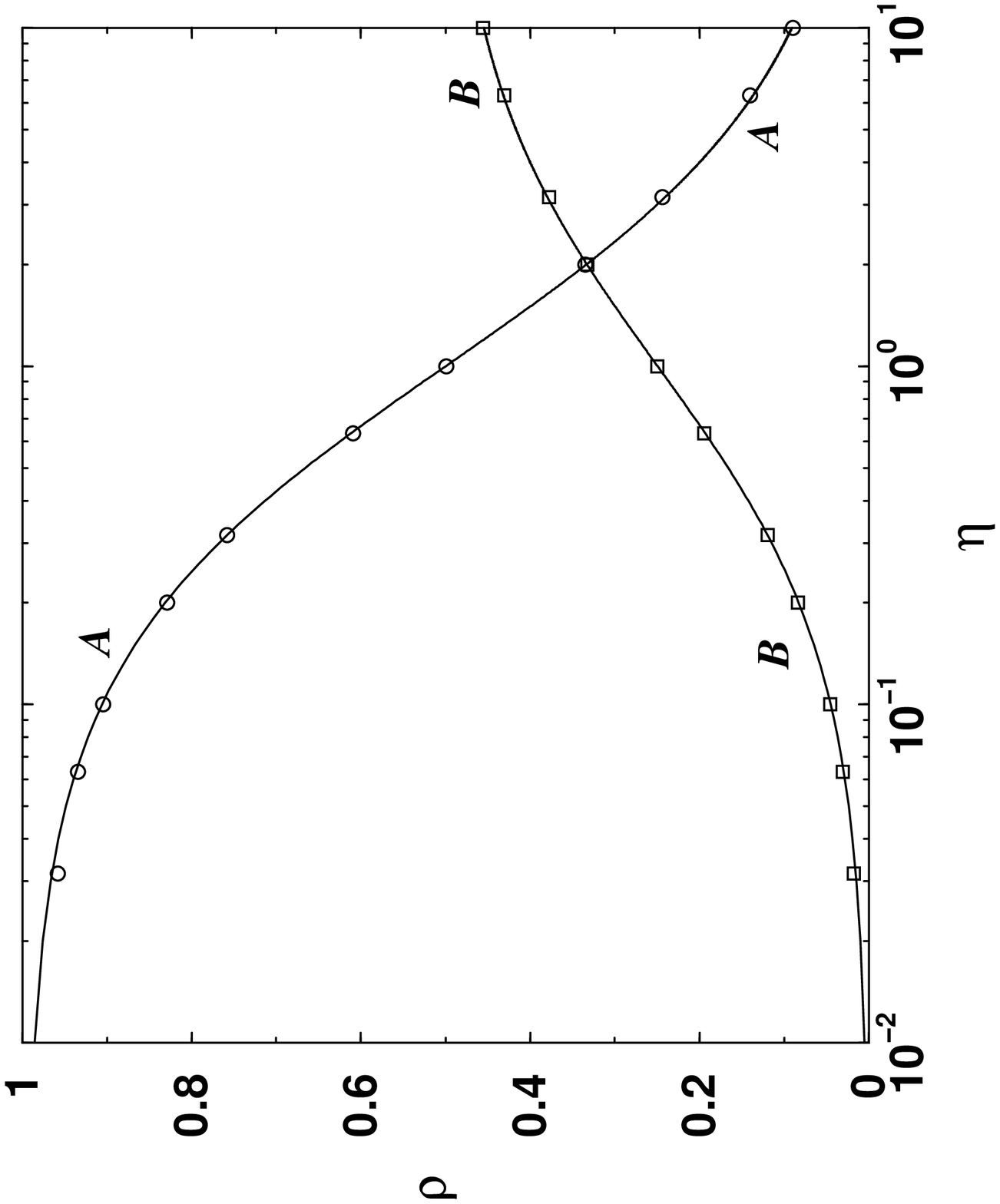}
 \caption{
Comparison between the numerical values of the steady densities, for particles
A (circles) and particles B (squares), obtained by Monte Carlo simulation, and
their  corresponding analytical expression (solid line), Eq. (\ref{15}), for
$\gamma_A=\gamma_B$. The agreement is quite good for $\gamma_A\leq 0.1$.}
 \label{f3}
 \caption{
Steady values of the densities of particles A (circles) and particles B
(squares), for $\gamma_A=0.01$, as a function of $\eta$. In the small $\eta$
region, $\rho_A\gg\rho_B$, while $\rho_B\gg\rho_A$ when $\eta$ is large.}
 \label{f4}
\end{figure}

In the weak vibration limit $\gamma_A\ll 1$, $\gamma_B\ll 1$ we are
considering, an asymptotic analysis, up to order $\gamma_A$, gives
\begin{equation}
    \rho_A = \frac{1}{1+\eta}-\frac{1}{2}\frac{1-\eta}{(1+\eta)^3}\gamma_A \, ,
    \quad
    \rho_B  =
    \frac{1}{2}\frac{\eta}{1+\eta}-\frac{1}{2}\frac{\eta}{(1+\eta)^3}\gamma_A
    \, ,
\label{15}
\end{equation}
uniformly valid for $\gamma_B\ll 1$ \cite{PyB03b}. In Eq.\ (\ref{15}), it is
$\eta=\gamma_A/\sqrt{\gamma_B}$.  Figure \ref{f3} compares the asymptotic
values of the densities, as given by Eq.\ \ref{15}, and the numerics, obtained
by Monte Carlo simulation, for $\gamma_A=\gamma_B$. The agreement is quite good
up to $\gamma_A \cong 0.1$. Also, in Fig. \ref{f4}, the asymptotics for the
densities is compared with the numerical values, for $\gamma_A=0.01$, as a
function of $\eta$. In this figure, it is clearly seen the crossover from BNE
to BNE of the model. In the region $\eta \ll 1$, we obtain $\rho_A\rightarrow
1$ and $\rho_B\rightarrow 0$, our system is full of small particles A (BNE). As
the size effect is taken into account by means of the different facilitation
rule, if both species are identical except for their size, it would be
$\alpha_a=\beta_a$ and $\alpha_d=\beta_d$, i. e., $\gamma_A=\gamma_B$ and
$\eta=\sqrt{\gamma_A}\ll 1$. This is why no RBNE is observed with particles of
the same material \cite{Wi76,RSPyS87}.

On the other hand, when $\eta \gg 1$, it is $\rho_A\rightarrow 0$ and
$\rho_B\rightarrow 1/2$. Now, the layer is full of big particles B. The line
separating these opposite behaviours can be determined by the condition
$\rho_A=\rho_B$. To the lowest order, this occurs for $\eta=2$, i.e.,
$\gamma_A=2\sqrt{\gamma_B}$. In order to understand why the parameter $\eta$
governs the transition from BNE to RBNE, let us consider the transitions
consisting in the substitution of  two particles A by one particle B. In Table
\ref{t1}, it is seen that the ratio of their effective rate $W_{\ab{ef}}$ to
those of their inverse transitions is $\eta^2$. In a low layer of a real
granular system, the desorption rates of the grains should decrease with their
mass density. In fact, a comparison with Refs. \cite{HQyL01,BEKyR03} indicates
that $\eta$ plays in the model a role analogous to the ratio of the mass
density of the big grains to the small ones in real granular systems and
hydrodynamic models. Therefore, the region $\eta \geq 2$, for which the RBNE is
expected, would correspond to large values of the mass density of the big
grains, as compared with the mass density of the small ones. Then, the picture
emerging from the model agrees with both numerical and experimental results
\cite{HQyL01,BEKyR03}. Interestingly, in the model, the RBNE is accompanied by
a non-trivial behaviour of the time evolution of the total density of particles
$\rho_A+\rho_B$, which presents a maximum at a certain time $t_c$. Roughly
speaking, the total density increases until the time $t_c$ for which
$\rho_A=\rho_B$; for greater times $\rho_B<\rho_A$ and the total density
decreases \cite{PyB03b}. We are not aware of experimental data allowing to
check this prediction.

In summary, a simple one-dimensional lattice model for an horizontal section of
a granular binary mixture, near the bottom of its container, has been
presented. The number of particles of this section of the system is not
conserved, when it is externally perturbed. This is a main difference with
other models for segregation, in which the number of particles are constant
\cite{NFyC02,Br02}. In our model, small (big) particles need one (both) of its
nearest neighbour sites empty in order to adsorb on or desorb from the lattice.
The system is submitted to a tapping process, so that in each free relaxation
the system evolves until it gets stuck in a metastable state. For weak
vibration, an effective dynamics connecting the metastable states has been
derived. The steady state distribution is consistent with Edwards' theory of
powders, and approximate analytical expressions for the steady densities have
been obtained. The steady distribution is characterized by the fugacities of
both species, which are related to their adsorption and desorption dynamical
rates. There are two well defined limit behaviours. Depending on the relative
values of the fugacities, the layer exhibits dominance of small particles (BNE)
or big particles (RBNE).

\acknowledgements

We acknowledge support from the Ministerio de Ciencia y Tecnolog\'{\i}a (Spain)
through Grant No.\ BFM2002-00307 (partially financed by FEDER funds).


\acknowledgments

\end{document}